\documentclass[aps,twocolumn,superscriptaddress,prd]{revtex4-1}   

\usepackage{amsmath}
\usepackage{graphicx}
\usepackage{epsfig}
\usepackage{dsfont}

\usepackage{bm}

\begin{document}

\title{A brief geometric derivation of some fundamental physics}

\author{Michel van Veenendaal}

\affiliation{Department of Physics, Northern Illinois University, DeKalb, Illinois 60115, USA}
\affiliation{Advanced Photon Source, Argonne National Laboratory, 9700 South Cass Avenue, Argonne, Illinois 60439, USA}

\date{\today}

\begin{abstract}
Several fundamental results in physics are derived  from the simple starting point of two commuting orthogonal unit vectors. The combination of these unit vectors leads to spherical harmonics and Schwinger's expression of the second-quantized angular momentum states in terms of bosonic operators. Commuting unit vectors can be turned into anticommuting ones by the restriction to a single unit vector. This leads to Pauli spin matrices. By including hyperbolic rotations, vectors can be given a finite norm and results from special relativity and Dirac's equation are found. The assumption that the change in four-momentum is due to the change in four-potential leads to the electromagnetic field tensor and the Lorentz force. Mawell's equations are obtained by viewing the four-potential as an harmonic oscillator driven by the four-current. The Schr\"odinger equation is obtained from the nonrelativistic limit of the four-momentum.
\end{abstract}

\maketitle
\section{Introduction}
Many problems in physics can be separated into a symmetry and a physical part. Consider \footnote{Obviously, many of the results here are known, but obtained from the perspective of a description in terms of vectors. More details can be found textbooks, such as Refs. \cite{Sakurai,GreinerRel,GreinerQED,Bjorken,Fitts,Lorrain,Griffiths}.}, as an example, a hydrogen atom as treated in many introductory quantum-mechanics textbooks \cite{Sakurai,GreinerRel,GreinerQED,Bjorken,Fitts,Lorrain,Griffiths}.
 The electron wavefunctions can be divided into a symmetry part, leading to the angular part of the wavefunction (given in spherical harmonics) and a physical part giving the radial part of the wavefunction (expressed in Laguerre polynomials). The symmetry part is entirely determined by the spherical symmetry of the atom, whereas the physical part depends on the details of the problem, namely the $1/r$ potential. This is not to say that the symmetry part does not contain any physics. On the contrary, many important aspects in physics are directly related to the symmetry of the problem. Symmetry in a particular space gives rise to conservation laws in its dual space. For the hydrogen atom, the spherical symmetry is intimately connected to the conservation of angular momentum. Therefore, when changing the physical aspect of the problem, the symmetry part of the solution remains unchanged, whereas the physical part is affected. For example, for a many-electron atom, the angular part of the problem is still given by the spherical harmonics, whereas the radial wavefunctions depend on the details of the system under consideration. Obviously, this division is not absolute and, for the hydrogen atom, one can also view the $1/r$ potential in terms of symmetry, leading to step operators for the radial wavefunction \cite{Fitts}.

There are often  a variety of approaches to solving the symmetry part of the problem. The spherical harmonics can be obtained by solving the angular part of the Laplacian or by using the step operators of the angular momentum operator. For other problems, different approaches are used to incorporate the symmetry of the problem. The Pauli  and $\gamma$ matrices are an essential ingredient to express the properties of vectors in space  and spacetime, respectively. However, it is also well known that there is a direct connection between angular momentum that arises in the hydrogen-atom problem and the Pauli spin matrices. In this paper, the premise  is that many fundamental results in physics are a direct result of the properties of vectors in different spaces. The starting point is a basis of  two commuting unit vectors ${\bm a}_\sigma^\dagger$ with $\sigma=\downarrow,\uparrow$. By combining these fundamental unit vectors, one can obtain commuting unit vectors ${\bm e}_i$ with $i=x,y,z$ in three dimensions. Note that this is essentially still a two dimensional system, since unit vectors are described by  two angular coordinates. Higher order combinations of the ${\bm a}_\sigma^\dagger$ directly lead to the spherical harmonics, without solving any differential equation. At the same time,  Schwinger's expression \cite{Schwinger} of the second-quantized form of the angular momentum states in  terms of bosonic operators given by the ${\bm a}_\sigma^\dagger$ is rederived. This is then followed by a crucial step where, by restricting oneself to a single ${\bm a}_\sigma^\dagger$, anticommuting unit vectors ${\bf e}_i$ are obtained. These unit vectors are directly related to the Pauli spin matrices. Up to that point, only unit vectors are considered, which can be rotated with the use of exponents of pseudovectors. Hyperbolic rotations or Lorentz boosts can be obtained by taking the exponent of a vector. These rotations can be used to give a finite norm to vectors, leading to four-vectors. This allows us to obtain some fundamental results from special relativity and the Dirac equation. From the assumption that the change in four-momentum is given by the change in four-potential, the electromagnetic field tensor, the power equation and the Lorentz force are derived. Maxwell's equations are obtained by viewing the four-potential as a driven harmonic oscillator with as driving force the four-current. The nonrelativistic limit of the four-momentum gives rise to the Schr\"odinger equation.

\section{Space: Commuting Unit Vectors}
\label{angmom}
A unit vector in a two-dimensional space can be written as
\begin{eqnarray}
{\hat {\bm r}}^\dagger=\cos\theta {\bm a}^\dagger_\uparrow+\sin\theta {\bm a}^\dagger_\downarrow,
\label{r2D}
\end{eqnarray}
where ${\bm a}_\sigma^\dagger$ with  $\sigma=\pm\frac{1}{2}=\uparrow,\downarrow$ creates a unit vector.
Its derivative with respect to $\theta$ is
\begin{figure}[t]
\begin{center}
\includegraphics[trim=0cm 6cm 13.8cm 3cm,width=0.95\columnwidth]{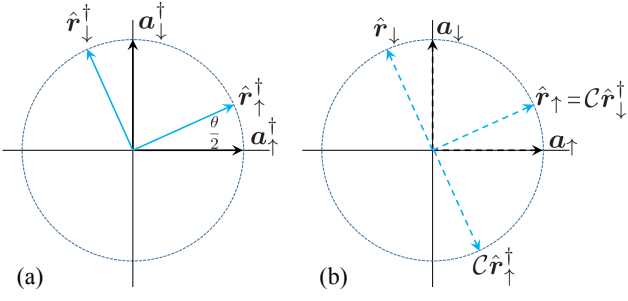}
\end{center}
\caption{\label{unitvectors}(a) The operators ${\bm a}^\dagger_\sigma$ with $\sigma=\uparrow,\downarrow$ create two orthogonal commuting unit vectors. The product ${\hat {\bm r}}_\uparrow^\dagger {\hat {\bm r}}_\downarrow^\dagger$ of the  half unit vectors ${\hat {\bm r}}_\sigma^\dagger$  defines a unit vector ${\hat {\bm r}}^\dagger$ in the $zx$ plane of a three-dimensional space. (b)  The annihilation  operators ${\bm a}_\sigma$ remove the unit vectors ${\bm a}^\dagger_\sigma$. ${\cal C}{\hat {\bm r}}_\sigma^\dagger$ shows the effect of the operator ${\cal C}={\bm a}_\uparrow{\bm a}_\downarrow -{\bm a}_\downarrow{\bm a}_\uparrow$ on the half unit vectors ${\hat {\bm r}}_\sigma^\dagger$. Note that the resulting vectors are, apart from a sign, related to the conjugates ${\hat {\bm r}}_\sigma$.
}
\end{figure}
\begin{eqnarray}
{\hat {\underline {\bm r}}}^\dagger= \frac{d{\hat {\bm r}}^\dagger}{d\theta}=-\sin\theta {\bm a}^\dagger_\uparrow+\cos\theta {\bm a}^\dagger_\downarrow.
\label{r2Dp}
\end{eqnarray}
This vector is perpendicular to the original vector. Instead of taking the derivative of the coefficients, the same operation can be achieved by manipulating the unit vectors. In this paper, when possible, the use of derivatives that lead to  differential equations is avoided and results are predominantly derived using operations on unit vectors.  A $90^\circ$ rotation is defined as
\begin{eqnarray}
{\hat {\underline {\bm r}}}^\dagger = {\bf i} {\hat {\bm r}}^\dagger,
\label{rotxz}
\end{eqnarray}
with
\begin{eqnarray}
{\bf i} = {\bm a}_\downarrow^\dagger {\bm a}_\uparrow-{\bm a}_\uparrow^\dagger {\bm a}_\downarrow.
\label{ixz12}
\end{eqnarray}
 For the moment, we can simply assume that ${\bm a}_\sigma$ annihilates ${\bm a}^\dagger_\sigma$, {\it i.e. } ${\bm a}_{\sigma'}{\bm a}^\dagger_\sigma =\delta_{\sigma\sigma'}$. The square of ${\bm i}$ can be written as
\begin{eqnarray}
{\bf i}^2 =-{\bm a}_\uparrow^\dagger {\bm a}_\uparrow-{\bm a}_\downarrow^\dagger {\bm a}_\downarrow.
\end{eqnarray}
${\bf i}^2$ leaves the unit vectors unchanged, but multiplies them by $-1$. The operator ${\bf i}$ therefore functions as the complex unit in a purely two-dimensional space.
Using this definition, we can also write a vector as
\begin{eqnarray}
{\hat {\bm r}}^\dagger = e^{{\bf i} \theta}  {\bm a}^\dagger_\uparrow.
\end{eqnarray}
The exponential is equivalent to a rotation in the plane
\begin{eqnarray}
R_\theta= e^{{\bf i} \theta}  =\cos\theta +\sin\theta {\bf i}.
\label{rot2D}
\end{eqnarray}
Two consecutive rotations are given by
\begin{eqnarray}
R_{\theta'}R_\theta= e^{{\bf i} \theta'}e^{{\bf i} \theta}  = e^{{\bf i} (\theta'+\theta)}=R_{\theta+\theta'}.
\nonumber
\end{eqnarray}

\begin{figure}[t]
\begin{center}s
\includegraphics[trim=2cm 12cm 12.5cm 0cm,width=0.95\columnwidth]{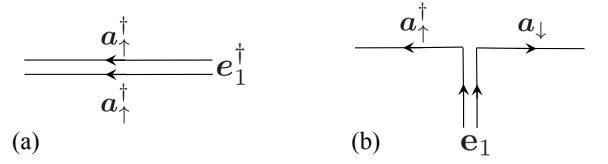}
\end{center}
\caption{\label{unit1} Schematic representation of commuting and anticommuting unit vectors. (a) For a commuting unit vector ${\bm e}^{\dagger}_m$ with $m=1,0,-1$ (${\bm e}^{\dagger}_1$ is shown as an example) a product ${\bm a}_\sigma^\dagger{\bm a}_{\sigma'}^\dagger$ of two unit vectors is created, where $m=\sigma+\sigma'$. The arrows for the spin should be read from right to left. (b) For an anticommuting unit vector ${\bf e}_m$, a unit vector is first annihilated with ${\bm a}_\sigma$ and a new unit vector is created with ${\bm a}_{\sigma'}^\dagger$  (${\bf e}_1$ is shown as an example); $m=\sigma'-\sigma$ gives the change between the two unit vectors.
}
\end{figure}

In physics, we are usually more interested in three-dimensional space. Let us first consider unit vectors in three dimensions. Since a unit vector has two degrees of freedom, it is  an effective two-dimensional space (note that unit vectors in two dimensions are an effective one-dimensional space with one coordinate).  By using pairs of two-dimensional unit vectors  ${\bm a}_{\sigma'}^\dagger{\bm a}_\sigma^\dagger$, new unit vectors can be created. In analogy to Eqs. (\ref{r2D}) and (\ref{r2Dp}), let us first define the half unit vectors
\begin{eqnarray}
{\hat {\bm r}}_\uparrow^\dagger=\cos\frac{\theta}{2} {\bm a}^\dagger_\uparrow+\sin\frac{\theta}{2} {\bm a}^\dagger_\downarrow, ~~~
{\hat {\bm r}}_\downarrow^\dagger=-\sin\frac{\theta}{2} {\bm a}^\dagger_\uparrow+\cos\frac{\theta}{2} {\bm a}^\dagger_\downarrow,
\end{eqnarray}
see Fig. \ref{unitvectors}.
Let us now consider the following product
\begin{eqnarray}
{\hat {\bm r}}_\uparrow^\dagger {\hat {\bm r}}_\downarrow^\dagger&=&(\cos^2\frac{\theta}{2}-
\sin^2\frac{\theta}{2}) {\bm a}^\dagger_\uparrow{\bm a}^\dagger_\downarrow  \nonumber \\
&& ~~~~+\sin\frac{\theta}{2}\cos \frac{\theta}{2}
(-{\bm a}^\dagger_\uparrow{\bm a}^\dagger_\uparrow+{\bm a}^\dagger_\downarrow{\bm a}^\dagger_\downarrow ).
\label{rr}
\end{eqnarray}
We have now three distinct combinations of the original unit vectors, which we use to define spherical unit vectors in three dimensions
\begin{eqnarray}
{\bm e}^{\dagger}_1= \frac{1}{\sqrt{2}}{\bm a}^\dagger_\uparrow{\bm a}^\dagger_\uparrow ,
~~{\bm e}^{\dagger}_0= {\bm a}^\dagger_\uparrow{\bm a}^\dagger_\downarrow , ~~
{\bm e}^{\dagger}_{-1}= \frac{1}{\sqrt{2}}{\bm a}^\dagger_\downarrow{\bm a}^\dagger_\downarrow .
\end{eqnarray}
A schematic depiction of these unit vectors is given in Fig. \ref{unit1}(a). The square root is introduced to keep the unit vectors normalized. The creation and annihilation operators satisfy the following property
\begin{eqnarray}
 {\bm a}_\sigma( {\bm a}^\dagger_\sigma)^m =m ( {\bm a}^\dagger_\sigma)^{m-1},
\end{eqnarray}
that is, there are $m$ different possibilities for $ {\bm a}_\sigma$ to annihilate a ${\bm a}^\dagger_\sigma$. The unit vectors ${\bm a}^\dagger_\sigma$ have the same properties as the step operators of a quantum harmonic oscillator. Note that this operation is similar to differentiation where
\begin{eqnarray}
\frac{d}{dx} x^m =m x^{m-1}.
\end{eqnarray}
Again, the manipulation of the unit vectors directly reflects the action of the derivative in the coordinates and vice-versa.
When two terms of the same power annihilate each other, we are left with the factor
\begin{eqnarray}
 ({\bm a}_\sigma)^m ( {\bm a}^\dagger_\sigma)^m =m! ~,
\end{eqnarray}
which gives a normalization factor of $1/\sqrt{m!}$. We want the quantity ${\hat {\bm r}}_\uparrow^\dagger {\hat {\bm r}}_\downarrow^\dagger$ again to represent a vector in a plane but now in a three dimensional space. Making use of the double-angle formulas we obtain for Eq. (\ref{rr})
\begin{eqnarray}
{\hat {\bm r}}_\uparrow^\dagger {\hat {\bm r}}_\downarrow^\dagger=\cos\theta {\bm e}_z^\dagger +\sin\theta {\bm e}_x^\dagger,
\label{rinter}
\end{eqnarray}
with
\begin{eqnarray}
{\bm e}_z^{\dagger}= {\bm e}^{\dagger}_0,~~~
{\bm e}_x^{\dagger}= \frac{1}{\sqrt{2}}(-{\bm e}^{\dagger}_1+{\bm e}^{\dagger}_{-1}).
\label{ezex}
\end{eqnarray}

We now want to introduce the third dimension using the additional degree of freedom obtained by combining ${\bm a}^\dagger_\sigma$. Let us first consider the operator
\begin{eqnarray}
{\bf i}_{xz}&=&\frac{\bf i}{2} =\frac{1}{2}( {\bm a}_\downarrow^\dagger {\bm a}_\uparrow-{\bm a}_\uparrow^\dagger {\bm a}_\downarrow)
\label{ixz1}
\end{eqnarray}
Above, we saw that ${\bf i}$ works as a rotation on the unit vectors in a two-dimensional plane ${\bm a}_\sigma$. By introducing the factor $\frac{1}{2}$, it still functions in the same way. After some work, we can show that
\begin{eqnarray}
{\bf i}_{xz}{\bm e}^\dagger_z={\bm e}^\dagger_x , ~~~{\rm and } ~~~ {\bf i}_{xz}{\bm e}^\dagger_x=-{\bm e}^\dagger_z,
\end{eqnarray}
which is what one would expect for an anticlockwise $90^\circ$ rotation in the $zx$ plane. We can also consider the diagonal operator
\begin{eqnarray}
{\bf i}_{yx} &=&\frac{i}{2}( {\bm a}_\uparrow^\dagger {\bm a}_\uparrow-{\bm a}_\downarrow^\dagger {\bm a}_\downarrow) ,
\label{iyx1}
\end{eqnarray}
where ${\bm a}_\sigma^\dagger {\bm a}_\sigma$ is the number operator. First, we must understand why the imaginary unit $i$ is necessary. If we had taken the operator ${\bf j}_{yx}=( {\bm a}_\uparrow^\dagger {\bm a}_\uparrow-{\bm a}_\downarrow^\dagger {\bm a}_\downarrow)/2$, subsequent operations starting from ${\bm e}^\dagger_x$, would give ${\bm e}^\dagger_x\rightarrow {\bf j}_{yx}{\bm e}^\dagger_x\rightarrow{\bm e}^\dagger_x\rightarrow\cdots$. This is not a rotation, but a hyperbolic transformation. The introduction of the factor $i$ gives ${\bm e}^\dagger_x\rightarrow{\bm e}^\dagger_y\rightarrow-{\bm e}^\dagger_x\rightarrow-{\bm e}^\dagger_y\rightarrow{\bm e}^\dagger_x\rightarrow\cdots$, where ${\bm e}^\dagger_y=- {\bf i}_{yx}{\bm e}^\dagger_x$.
 The minus sign in front of ${\bf i}_{yx}$ is introduced since we are looking at the conjugate of the vectors. This sequence is an anticlockwise $90^\circ$ rotation. The additional unit vector is therefore
\begin{eqnarray}
{\bm e}_y^{\dagger}=-{\bf i}_{yx}{\bm e}_x^{\dagger}=\frac{i}{\sqrt{2}}({\bm e}^{\dagger}_1+{\bm e}^{\dagger}_{-1}),
\label{ey}
\end{eqnarray}
giving the third unit vector.  Additionally, we have  ${\bf i}_{yx}{\bf e}_z=0$. Operating with ${\bf i}_{yz} $ on the spherical unit vectors gives
\begin{eqnarray}
{\bf i}_{yx} {\bm a}^\dagger_\sigma=i\sigma  {\bm a}^\dagger_\sigma ~~~{\rm and }~~~{\bf i}_{yx} {\bm e}^{\dagger}_m=i m {\bm e}^{\dagger}_m,
\end{eqnarray}
where $\sigma=\pm \frac{1}{2}$ and $m=1,0,-1$. By forcing the rotation of ${\bm e}^{\dagger}_i$ with $i=x,y$ to be unitary, the eigenvalues of ${\bm a}^\dagger_\sigma$  have become half integers. This shows that ${\bf i}_{yx}$ performs the same operation as the total angular momentum operator $i{\rm J}_z$. A three-dimensional unit vector can be obtained by rotating the $x$-axis in Eq. (\ref{rinter})  using  $ {\bm e}^\dagger _{x'}=e^{-{\bf i}_{yx} \varphi} {\bm e}^\dagger _x$. This gives
\begin{eqnarray}
{\hat {\bm r}}^\dagger&=& {\hat {\bm r}}_\uparrow^\dagger {\hat {\bm r}}_\downarrow^\dagger=\cos\theta {\bm e}_z^\dagger +\sin\theta
 {\bm e}_{x'}^\dagger \nonumber \\
&=&\cos\theta {\bm e}_z^\dagger +\sin\theta \cos\varphi {\bm e}_x^\dagger +\sin\theta \sin\varphi {\bm e}_y^\dagger,
\end{eqnarray}
which is a unit vector in spherical polar coordinates.

Since  unit vectors in three-dimensions occupy an effective two-dimensional space (given by the two angular coordinates), they can also be built by combinations of  the following half unit vectors:
\begin{eqnarray}
{\hat {\bm r}}_\uparrow^\dagger&=&\cos\frac{\theta}{2} e^{ -{\bf i}_{yx} \frac{\varphi}{2}}{\bm a}^\dagger_\uparrow+\sin\frac{\theta}{2}e^{ {\bf i}_{yx} \frac{\varphi}{2}} {\bm a}^\dagger_\downarrow,  \nonumber \\
{\hat {\bm r}}_\downarrow^\dagger&=&-\sin\frac{\theta}{2}e^{ -{\bf i}_{yx} \frac{\varphi}{2}}{\bm a}^\dagger_\uparrow+\cos\frac{\theta}{2}e^{ {\bf i}_{yx} \frac{\varphi}{2}} {\bm a}^\dagger_\downarrow.
\nonumber
\end{eqnarray}
 The half unit vectors ${\hat {\bm r}}_\uparrow^\dagger$ and ${\hat {\bm r}}_\downarrow^\dagger$  are essentially  Pauli spinors. We can now rewrite Eq. (\ref{rr}):
\begin{eqnarray}
&&{\hat {\bm r}}^\dagger ={\cal R}[{\hat {\bm r}}_\uparrow^\dagger {\hat {\bm r}}_\downarrow^\dagger]=\cos \theta
{\bm e}_0^\dagger \nonumber \\
&& ~~~~~~~~+\left (- \frac{\sin\theta}{\sqrt{2}} e^{-{\bf i}_{yx} \varphi} \right )
{\bm e}_1^\dagger + \left (\frac{\sin\theta}{\sqrt{2}} e^{{\bf i}_{yx} \varphi} \right ) {\bm e}_{-1}^\dagger  . \nonumber
\end{eqnarray}
The reordering operator ${\cal R}$ is introduced. In general, the rotations over different axes do not commute with each other. In this case, ${\bf i}_{yx}$ and ${\bm a}^\dagger_\sigma$ do not commute. ${\cal R}$ reorders all rotations so that first all the rotations in the $zx$ plane are performed, followed by  the rotations around the $z$ axis. Alternatively, one can maintain the order of the rotations, but define all the rotations over $\varphi$ around the original $z$ axis and the rotations over $\theta$ in the rotated $zx$ plane.  In the coefficients, we recognize the spherical harmonics. The renormalized spherical harmonics $C^l_m$ are related to the usual spherical harmonics by \cite{Varshalovich}
\begin{eqnarray}
C^l_m ({\hat {\bf r}}) =\sqrt{\frac{4\pi}{2l+1}} Y_{lm}({\hat {\bf r}}),
\end{eqnarray}
using ${\hat {\bf r}}$ as a shorthand for the angular coefficients $\theta$ and $\varphi$. The conjugate of the spherical harmonic is
\begin{eqnarray}
C^{lm}({\hat {\bf r}})\equiv \left (C^{l}_m({\hat {\bf r}}) \right )^*=
(-1)^m C^l_{-m} ({\hat {\bf r}})
\end{eqnarray}
The unit vector can then be written as
\begin{eqnarray}
{\hat {\bm r}}^\dagger =\sum_{m=-1}^1 C^{1m}({\hat {\bf r}}) {\bm e}^\dagger_m .
\label{r1}
\end{eqnarray}

We can extend this procedure to arbitrary powers, which gives
\begin{eqnarray}
{\hat {\bm r}}^{(l)\dagger} &=&{\cal R}\left [\frac{1}{l!}({\hat {\bm r}}_\uparrow^\dagger {\hat {\bm r}}_\downarrow^\dagger)^l \right ]  =
\sum_{m=-l}^l C^{lm}({\hat {\bm r}}) {\bm e}^{l\dagger}_m
\label{rexpand}
\end{eqnarray}
For rank 1, we use the convention ${\hat {\bm r}}^{\dagger}\equiv{\hat {\bm r}}^{(1)\dagger}$ and ${\bm e}^{\dagger}_m\equiv {\bm e}^{1\dagger}_m$, thereby dropping the rank $l$ when its equal to 1.
The unit vectors are expressed in terms of ${\bm a}^\dagger_\sigma$ as
\begin{eqnarray}
{\bm e}^{l\dagger}_m\equiv |lm\rangle  =\frac{ ({\bm a}^\dagger_\uparrow)^{l+m} ({\bm a}^\dagger_\downarrow)^{l-m} }{\sqrt{(l+m)!(l-m)!} },
\label{elm}
\end{eqnarray}
which rederives Schwinger's expression for the angular momentum basis.  The unit vector ${\bm e}^{l\dagger}_m$ can also be denoted  as $|lm\rangle $ in Dirac's bra-ket notation.  In the Schwinger basis, the angular momentum operators are given by  \cite{Judd,Brink}
\begin{eqnarray}
{\rm L}_0&=&-i\,{\bf i}_{yx}=\frac{1}{2}({\bm a}^\dagger_\uparrow{\bm a}_\uparrow-{\bm a}^\dagger_\downarrow{\bm a}_\downarrow)\\
{\rm L}_+ &=&{\bm a}^\dagger_\uparrow {\bm a}_\downarrow, ~~~~~~{\rm L}_- ={\bm a}^\dagger_\downarrow {\bm a}_\uparrow,
\end{eqnarray}
taking $\hbar\equiv 1$. These expressions are significantly simpler than those expressed in terms of derivatives, showing the advantage of manipulating unit vectors as opposed to coordinates. The $z$ component ($m=0$) gives ${\rm L}_0|lm\rangle=m |lm\rangle$.  ${\rm L}_\pm$ are the step operators that raise or lower the value of $m$:
\begin{eqnarray}
{\rm L}_\pm |lm\rangle= \sqrt{(l\mp m)(l\pm m+1)} |l,m\pm 1\rangle,
\end{eqnarray}
which can be evaluated using the expressions for the step operators for the quantum harmonic oscillator ($a|n\rangle=\sqrt{n}|n-1\rangle$ and $a^\dagger|n\rangle=\sqrt{n+1}|n+1\rangle$) . In bra-ket notation, Equation (\ref{rexpand}) is written as
\begin{eqnarray}
|{\hat {\bm r}}^{(l)}) &=&\sum_{m=-l}^l |lm\rangle \langle lm |{\hat {\bm r}}),
\end{eqnarray}
with $C^{lm}({\hat {\bf r}} )=\langle lm |{\hat {\bf r}}) $.  A parenthesis is used in $|{\hat {\bf r}}) $ to indicate that $C^{lm}({\hat {\bf r}} )$ is not orthonormal when integrating the angular coordinates \cite{Varshalovich}. The conjugate is given by  $C^{l}_m({\hat {\bf r}} )=({\hat {\bf r}}|lm\rangle $.

The renormalized spherical harmonics are equivalent to the spherical harmonics, except for a different starting point. For spherical harmonics, the starting point is the function space.  They are usually derived by solving the angular part of the Laplacian in spherical symmetry.  The emphasis is on the orthonormality of the wavefunctions. In bra-ket notation
\begin{eqnarray}
\langle l'm'|lm\rangle &=&\int d{\hat {\bf r}} \langle l'm'|{\hat {\bf r}}\rangle \langle {\hat {\bf r}}|lm\rangle
\nonumber \\ &=&
\int d{\hat {\bf r}} \,  Y_{l'm'}^*({\hat {\bf r}} )Y_{lm}({\hat {\bf r}} )=\delta_{ll'}\delta_{mm'},
\end{eqnarray}
with $ Y_{lm}({\hat {\bf r}} )=\langle {\hat {\bf r}}|lm\rangle$ which is the projection of the function $|lm\rangle$ on the unit vector  ${\hat {\bf r}}$.
For renormalized spherical harmonics, the starting point is the vector space and the coordinates are obtained by a proper manipulation of the unit vectors.
The focus is therefore on the orthonormality of vectors
\begin{eqnarray}
({\hat {\bf r}}'^{(l)}|{\hat {\bf r}}^{(l)})&=&\sum_m ({\hat {\bf r}}'|lm\rangle\langle lm |{\hat {\bf r}})
\nonumber \\ &=&
\sum_m C^l_m({\hat {\bf r}}' )[C^l_m({\hat {\bf r}})]^* =\delta({\hat {\bf r}}-{\hat {\bf r}}').
\end{eqnarray}
This expression, apart from a factor, is the completeness relation for the spherical harmonics.

An expression for the spherical harmonics can be obtained by using a binomial expansion in Eq. (\ref{rexpand}). For the conjugate ${\hat {\bm r}}^{(l)}$, we find
\begin{eqnarray}
&&{\hat {\bm r}}^{(l)} =\frac{1}{l!}{\cal R}\left [\sum_{\mu=0}^l  \left (
\begin{array}{c}
l \\ \mu
\end{array}
\right ) \left  (\cos \frac{\theta}{2}{\bm a}_\uparrow  \right )^\mu
\left  (\sin \frac{\theta}{2}{\bm a}_\downarrow  \right )^{l-\mu}  \right .\nonumber \\
&& ~~~~\times \left .
\sum_{\mu'=0}^l  \left (
\begin{array}{c}
l \\ \mu'
\end{array}
\right )
\left  (-\sin \frac{\theta}{2}{\bm a}_\uparrow  \right )^{l-\mu'}
\left  (\cos \frac{\theta}{2}{\bm a}_\downarrow  \right )^{\mu'}  \right ]. \nonumber
\end{eqnarray}
From Eq. (\ref{elm}), we see that $C^l_m({\hat {\bf r}})$ is determined by all the terms for which ${\bm a}_\uparrow$ is raised to the power $l+m$. This gives the condition  $\mu+l-\mu'=l+m$ or $\mu'=m-\mu$. This gives for the renormalized spherical harmonic
\begin{eqnarray}
&&C^l_m({\hat {\bf r}})=\frac{1}{l!}\sqrt{(l+m)!(l-m)!}  \sum_{\mu=0}^l  \left (
\begin{array}{c}
l \\ \mu
\end{array}
\right )
\left (
\begin{array}{c}
l \\ \mu-m
\end{array}
\right ) \nonumber \\
&&~~\times
(-1)^{l+m-\mu}\left (
\cos \frac{\theta}{2} \right )^{2\mu-m}
\left (
\sin \frac{\theta}{2} \right )^{2l+m-2\mu} e^{{\bf i}_{yx}m \varphi} \nonumber
\end{eqnarray}

Since the half unit vectors are only part of a unit vector, they behave differently from what one would expect for a vector. For example, under   a time-reversal transformation, a vector is inverted, {\it i.e.} $T{\hat {\bm r}}^\dagger=-{\hat {\bm r}}^\dagger$, where $T$ is the time-reversal operator. For the half unit vectors, we can write this as $T{\hat {\bm r}}^\dagger=-{\hat {\bm r}}_\uparrow^\dagger {\hat {\bm r}}_\downarrow^\dagger=({\bf i}{\hat {\bm r}}_\uparrow^\dagger)({\bf i }{\hat {\bm r}}_\downarrow^\dagger)$. Therefore, whereas the vector is inverted (or rotated by $180^\circ$), the half unit vectors are only rotated by $90^\circ$. For a general unit vector, we find for the time reveral
\begin{eqnarray}
T |jm\rangle  =\frac{ ({\bm a}^\dagger_\downarrow)^{j+m} (-{\bm a}^\dagger_\uparrow)^{j-m} }{\sqrt{(j+m)!(j-m)!} }
=(-1)^{j-m} |j,-m\rangle,
\end{eqnarray}
where $j$ indicates that half-integer values are allowed. For real vectors, this operation produces the same result as conjugation. For half-integer $j$ values, applying the time-reversal operator twice does not return the original state
\begin{eqnarray}
T ^2|jm\rangle
=(-1)^{j-m}T |j,-m\rangle=(-1)^{2j} |jm\rangle,
\end{eqnarray}
whereas conjugation does
\begin{eqnarray}
(|jm\rangle)^{\dagger \dagger}=(\langle jm |)^\dagger= |jm\rangle,
\end{eqnarray}
which is a direct result of $({\bm a}_\sigma^\dagger)^\dagger ={\bm a}_\sigma $.

\section{Space: Anticommuting Unit Vectors}

Unit vectors are described by the product ${\hat {\bm r}}^\dagger ={\hat {\bm r}}_\uparrow^\dagger {\hat {\bm r}}_\downarrow^\dagger$, implicitly assuming the reordering operator ${\cal R}$. However, all the information of the unit vector is already contained in the half unit vectors ${\hat {\bm r}}_\sigma^\dagger$. An important vector in physics is the momentum ${\bm p}$ whose direction is given by the unit vector ${\hat {\bm p}}$. If we want to study change in the direction of the momentum, it suffices to only look at changes in the half unit vectors or spinors ${\hat {\bm p}}_\sigma^\dagger$.
To study these changes, it is convenient to introduce a new set of unit vectors. Let us define the  operator
\begin{eqnarray}
{\cal C}={\bm a}_\uparrow{\bm a}_\downarrow -{\bm a}_\downarrow{\bm a}_\uparrow ,
\end{eqnarray}
which is a rotation over $-90^\circ$, but with the ${\bm a}^\dagger_\sigma$ operators replaced by their conjugates ${\bm a}_\sigma$.
New unit vectors can then be defined as
\begin{eqnarray}
{\bf e}_q={\cal C}{\bm e}_q^\dagger.
\end{eqnarray}
Note also the different font. These unit vectors are given by
\begin{eqnarray}
{\bf e}_1&=&-\sqrt{2} {\bm a}^\dagger_\uparrow {\bm a}_\downarrow \nonumber \\
{\bf e}_0 &=& {\bm a}^\dagger_\uparrow{\bm a}_\uparrow-{\bm a}^\dagger_\downarrow{\bm a}_\downarrow \\
{\bf e}_{-1}&=&\sqrt{2} {\bm a}^\dagger_\downarrow {\bm a}_\uparrow .\nonumber
\end{eqnarray}
A schematic depiction of  the unit vectors ${\bf e}_m$, with $m=1,0,-1$  denoting the change in a single $\sigma$, is given in Fig. \ref{unit1}(b).
Using Eqs. (\ref{ezex}) and (\ref{ey}), we can also obtain the Cartesian unit vectors:
\begin{eqnarray}
{\bf e}_x&=&{\bm a}^\dagger_\uparrow {\bm a}_\downarrow +{\bm a}^\dagger_\downarrow {\bm a}_\uparrow \nonumber \\
{\bf e}_y&=&i(-{\bm a}^\dagger_\uparrow {\bm a}_\downarrow +{\bm a}^\dagger_\downarrow {\bm a}_\uparrow) \label{basisei}\\
{\bf e}_z &=& {\bm a}^\dagger_\uparrow{\bm a}_\uparrow-{\bm a}^\dagger_\downarrow{\bm a}_\downarrow .\nonumber
\end{eqnarray}
In the case where the is only one half unit vector ${\bm a}^\dagger_\sigma$ at the time, the Cartesian unit vectors can be written in matrix form
\begin{eqnarray}
{\rm e}_x=
\left (
\begin{array}{cc}
0 & 1 \\
1 & 0
\end{array}
\right ), ~~
{\rm e}_y=
\left (
\begin{array}{cc}
0 & -i \\
i & 0
\end{array}
\right ), ~~
{\rm e}_z=
\left (
\begin{array}{cc}
1 & 0 \\
0 & -1
\end{array}
\right ),
\label{Paulispin}
\end{eqnarray}
where nonitalic letters are used to indicate matrices. The basis of the matrix consists of the unit vectors $|\uparrow\rangle, |\downarrow\rangle$,  in bra-ket notation. The half unit vector can be written in this basis as  $|{\hat {\bm r}}_\uparrow\rangle=a|\uparrow\rangle+b|\downarrow\rangle$. These unit matrices are generally known as the Pauli spin matrices with $\sigma_i\equiv {\rm e}_i$. Instead of expression a vector in the unit vectors ${\bf e}_i=\sigma_i$ with $i=x,y,z$, we can also express a vector in terms of unit matrices ${\rm e}_i=\sigma_i$
\begin{eqnarray}
{\rm r}=\sum_{i=x,y,z} r_i {\rm e}_i.
\end{eqnarray}
Note that the summation on  the right-hand side is often written as $({\bf r}\cdot {\bm \sigma})$. This notation has the disadvantage that it appears as a scalar (an inner product between two vectors), whereas the quantity is actually a vector.  Interpreting the Pauli spin matrices as effective unit ``vectors" is a key ingredient in geometric algebra \cite{Hestenes,Doran}.

\begin{figure}[t]
\begin{center}
\includegraphics[trim=0cm 9cm 15cm 2cm,width=1.06\columnwidth]{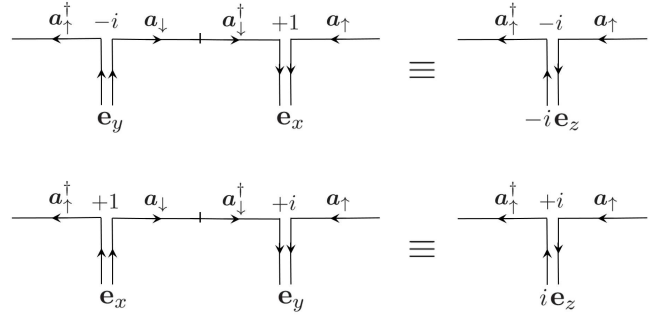}
\end{center}
\caption{\label{unitcommuting} Schematic representation of the anticommuting nature of the unit vectors ${\bf e}_i$ is related to the different result obtained when changing the order of the unit vectors working on a single ${\bm a}^\dagger_\sigma$. The figure shows the example of ${\bf e}_x$ and ${\bf e}_y$ working on ${\bm a}^\dagger_\uparrow$. The diagram should be read from right to left.  The combined action of the unit vectors ${\bf e}_x$ and ${\bf e}_y$ can also be expressed in terms of ${\bf e}_z$.
}
\end{figure}

It is straightforward to show that the norm of the unit vectors is given by ${\bf e}_i^2=1$, Likewise, the Pauli spin matrices satisfy ${\rm e}_i^2={\mathds 1}_2$, where $={\mathds 1}_n$ is a $n\times n$ identity matrix. However, something interesting has happened. The original unit vectors commuted with each other, {\it i.e.} ${\bm e}_i{\bm e}_j={\bm e}_j{\bm e}_i$. However, when there is a restriction to a single ${\bm a}^\dagger_\sigma$, the order becomes important. In that case, the new unit vectors follow the anticommutation relation
\begin{eqnarray}
{\bf e}_i{\bf e}_j +{\bf e}_j{\bf e}_i =2 \delta_{ij} ~~~{\rm for }~~~ i,j=x,y,z.
\end{eqnarray}
The anticommuting nature of the unit vectors is expressed schematically in Fig. \ref{unitcommuting}. The Pauli spin matrices satisfy the same relations. The combination of two vectors is called a  bivector
\begin{eqnarray}
{\bf i}_{ij}={\bf e}_i{\bf e}_j ~~~{\rm for }~~~ i\ne j
\end{eqnarray}
The bivectors ${\bf i}_{xz}$ and ${\bf i}_{yx}$ differ by a factor 2 from those in Eqs. (\ref{ixz1}) and (\ref{iyx1}), respectively. However, ${\bf i}_{xz}$ is equivalent to ${\bf i}$ in Eq. (\ref{ixz12}). The reason for this is that equations (\ref{ixz1}) and (\ref{iyx1}) refer to the basis vectors ${\bm e}_i$, which consist of ${\bm a}^\dagger_\sigma {\bm a}^\dagger_{\sigma'}$ pairs, whereas the current bivectors only operate on a single basis vector ${\bm a}^\dagger_\sigma $.
A bivector is a complex unit since
\begin{eqnarray}
{\bf i}_{ij}^2={\bf e}_i{\bf e}_j{\bf e}_i{\bf e}_j=- {\bf e}_i{\bf e}_i{\bf e}_j{\bf e}_j=-1,
\end{eqnarray}
using the commutation relation.
 Additionally, in three dimensions, there is a single trivector
\begin{eqnarray}
{\bf i}={\bf e}_x{\bf e}_y{\bf e}_z, ~~~{\rm with }~~~{\bf i}^2=-1.
\end{eqnarray}
The notation ${\bf i}$ is also used in Eq. (\ref{ixz12}).  Both quantities are pseudoscalars, {\it i.e.} the highest grade element in a particular dimension.
In terms of Pauli spin matrices, this quantity is ${\rm e}_x{\rm e}_y{\rm e}_z=i {\mathds 1}_2$. The trivector allows us to write bivectors as a (pseudo) vector ${\underline {\bf e}}_k$
\begin{eqnarray}
{\underline {\bf e}}_k={\bf i}{\bf e}_k= {\bf i}_{ij}={\bf e}_i{\bf e}_j ,
\label{pseudovector}
\end{eqnarray}
with $k=x,y,z$ for $ij=yz,zx,xy$. Note that, whereas in three dimensions, the pseudovector is a bivector, in two dimensions pseudovectors and vectors are equivalent since  ${\bf e}_y={\bf i}{\bf e}_x$, as we saw in Eq. (\ref{rotxz}).

The definition of the pseudovector also provides a convenient way to write rotations. By defining the angle pseudovector
\begin{eqnarray}
{\underline {\bm \varphi}}=\varphi {\underline {\bf e}_z}= \varphi {\bf i}{\bf e}_z=\varphi {\bf e}_x {\bf e}_y={\bf i}{\bm \varphi},
\end{eqnarray}
with the angle vector ${\bm \varphi}= \varphi{\bf e}_z$,
a rotation around the $z$ axis can be written as
\begin{eqnarray}
e^{-{\underline {\bm \varphi}}}&=&\cosh {\underline {\bm \varphi}}-\sinh {\underline {\bm \varphi}}
=\cos {\bm \varphi} -{\bf i} \sin {\bm \varphi}\label{rotation}
 \\
&=&\cos \varphi -{\bf i}{\bf e}_z \sin \varphi =\cos \varphi +{\bf e}_y{\bf e}_x \sin \varphi=e^{{\bf i}_{yx}\varphi}.
  \nonumber
\end{eqnarray}

A vector in the ${\bf e}_i$ basis can be written as
\begin{eqnarray}
{\bf r}=\sum_{i=x,y,z} r_i {\bf e}_i
=x{\bf e}_x+y{\bf e}_y+z{\bf e}_z.
\end{eqnarray}
This vector can also be subdivided into half unit vectors
\begin{eqnarray}
{\bf r}={\cal C}({\hat {\bm r}}_\uparrow^\dagger{\hat {\bm r}}_\downarrow^\dagger)
={\hat {\bm r}}_\uparrow^\dagger{\cal C}{\hat {\bm r}}_\downarrow^\dagger+{\hat {\bm r}}_\downarrow^\dagger{\cal C}{\hat {\bm r}}_\uparrow^\dagger
\end{eqnarray}
where the rotation reordering ${\cal R}$ has been assumed implicitly. The conjugation of ${\hat {\bm r}}_\sigma^\dagger$ can be written as
\begin{eqnarray}
{\hat {\bm r}}_\uparrow={\cal C}{\hat {\bm r}}_\downarrow^\dagger=
&=&\cos\frac{\theta}{2} e^{ {\bf i}_{yx} \frac{\varphi}{2}}{\bm a}_\uparrow+\sin\frac{\theta}{2}e^{ -{\bf i}_{yx} \frac{\varphi}{2}} {\bm a}_\downarrow,  \nonumber \\
{\hat {\bm r}}_\downarrow=-{\cal C}{\hat {\bm r}}_\uparrow^\dagger&=&-\sin\frac{\theta}{2}e^{ {\bf i}_{yx} \frac{\varphi}{2}}{\bm a}_\uparrow+\cos\frac{\theta}{2}e^{ -{\bf i}_{yx} \frac{\varphi}{2}} {\bm a}_\downarrow.
\nonumber
\end{eqnarray}
This operation is shown in Fig. \ref{unitvectors}(b). The unit vector ${\hat {\bf r}}$ can then be expressed in terms of half unit vectors or spinors as
\begin{eqnarray}
{\hat {\bf r}}
={\hat {\bm r}}_\uparrow^\dagger{\hat {\bm r}}_\uparrow-{\hat {\bm r}}_\downarrow^\dagger{\hat {\bm r}}_\downarrow.
\label{unitr}
\end{eqnarray}
The operation of ${\hat {\bf r}}$ on the spinors gives
\begin{eqnarray}
{\hat {\bf r}}{\hat {\bm r}}_{\pm \frac{1}{2}}^\dagger=\pm {\hat {\bm r}}_{\pm \frac{1}{2}}^\dagger.
\end{eqnarray}

Since the dimension of the space for the unit vectors is two  (given by the coordinates $\theta$ and $\varphi$), the spinors ${\hat {\bm r}}_\sigma^\dagger$ are sufficient to describe all  unit vectors. However, often it is convenient to describe the unit vectors in a three-dimensional basis. This can be done via
\begin{eqnarray}
{\hat {\bf r}}(\sigma) =\sum_{i=x,y,z}({\hat {\bm r}}_\sigma {\bf e}_i {\hat {\bm r}}_\sigma^\dagger) {\bm e}_i,
\end{eqnarray}
where ${\bm e}_i$ are the conjugates of the unit vectors given in Eqs. (\ref{ezex}) and (\ref{ey}). The term between parentheses is a scalar, namely the expectation value of the unit vectors from Eq. (\ref{basisei}) between the spinors ${\hat {\bm r}}_\sigma$ and ${\hat {\bm r}}_\sigma^\dagger$. In bra-ket notation, it can be written as ${\hat {\bm r}}_\sigma {\bf e}_i {\hat {\bm r}}_\sigma^\dagger\equiv \langle {\hat {\bm r}}_\sigma |{\rm e}_i|{\hat {\bm r}}_\sigma \rangle$.   The two spinors ${\hat {\bm r}}_\sigma^\dagger$ correspond to unit vectors in the opposite direction, {\it i.e.} ${\hat {\bf r}}(-\frac{1}{2})=-{\hat {\bf r}}(\frac{1}{2})$.

Finally, an important identity that we will need in the remainder is the product of two vectors
\begin{eqnarray}
{\bf a}{\bf b}={\bf a}\cdot {\bf b}+{\bf a}\wedge{\bf b}={\bf a}\cdot {\bf b}+{\bf i}\, {\bf a}\times {\bf b}.
\label{vecprod}
\end{eqnarray}
The product can be split into an inner product,
The product can be split into an inner product,
\begin{eqnarray}
{\bf a}\cdot {\bf b} =\sum_i a_i b_i,
\end{eqnarray}
 and a wedge product,
\begin{eqnarray}
{\bf a}\wedge {\bf b} =\sum_k ({\bf a}\times {\bf b} )_k {\underline {\bf e}}_k,
\end{eqnarray}
using the definition of the pseudovector from Eq. (\ref{pseudovector}).  The wedge product, which is a pseudovector or bivector, can be written a ${\bf i}$ times the outer product, which is a vector.  For two equal vectors the product reduces to the norm of the vector  ${\bf a}{\bf a}={\bf a}\cdot {\bf a}=a^2$.

\section{Spacetime}
In the preceding Section, we have mainly dealt with unit vectors. Equation (\ref{rotation}) shows that the rotation over an angle pseudovector corresponds to a rotation. Let us now look at the exponent of an angle vector
\begin{eqnarray}
{\bm \alpha}=\alpha {\hat {\bf p}},
\end{eqnarray}
where ${\hat {\bf p}}\equiv {\hat {\bf p}}(\frac{1}{2})$ is a unit vector in the direction of the momentum, {\it i.e.} the positive spin direction. To connect more closely to the physical properties of particles, we focus on the unit vector of the momentum.
The rotation is then
\begin{eqnarray}
R_{{\bm \alpha}\uparrow}=e^{\bm \alpha}=\cosh{\bm \alpha}+\sinh {\bm \alpha}=\cosh \alpha+{\hat {\bf p}} \sinh\alpha.
\end{eqnarray}
This quantity splits into a scalar $\cosh \alpha$  and a vector with a magnitude $ \sinh\alpha$. In special relativity, the scalar term is associated with a unit vector in time/energy ${\bf e}_0$, with ${\bf e}_0^2=1$ and ${\bf e}_0{\hat {\bf p}} =-{\hat {\bf p}}{\bf e}_0$. This creates an effective two-dimensional space spanned by the orthogonal unit vectors ${\bf e}_0$ and ${\hat {\bf p}}$. The four-momentum can then  be written as
\begin{eqnarray}
\textsf{\textbf{p}}=mc e^{\bm \alpha}{\bf e}_0=mc (\cosh \alpha {\bf e}_0+\sinh\alpha {\hat {\bf p}}' ),
\label{fourmomentum}
\end{eqnarray}
where four-vectors are indicated by sans-serif letters.
The  factor $mc$, with $m$ the mass and $c$ the speed of light, is introduced to obtain the proper unit and magnitude of the four-momentum.   The unit vector for the momentum  has now become a bivector
\begin{eqnarray}
{\hat {\bf p}}' ={\hat {\bf p}}{\bf e}_0 , ~~~{\rm with }~~~ {\hat {\bf p}}'^2 =-1.
\end{eqnarray}
 This produces a switch to the Minkowski metric \cite{Minkowski}.
The norm of the four-momentum then becomes
\begin{eqnarray}
\textsf{\textbf{p}}^2=m^2c^2  (\cosh^2 \alpha - \sinh^2\alpha)=m^2c^2,
\end{eqnarray}
and is therefore constant. The components in the ${\bf e}_0$ and ${\hat {\bf p}}' $ direction are identified as the energy $E$ and the momentum  ${\bf p}'$, respectively,
\begin{eqnarray}
\frac{E}{c}=mc \cosh \alpha ~~~{\rm and }~~~ {\bf p}'=mc \sinh\alpha\, {\hat {\bf p}}'.
\end{eqnarray}
 The $\cosh\alpha$ term is also known as  the Lorentz factor $\gamma$
\begin{eqnarray}
\gamma=\cosh\alpha =\frac{E}{mc^2}~~~{\rm or }~~~ E=\gamma mc^2,
\end{eqnarray}
which is Einstein's mass-energy equivalence formula \cite{Einstein}. The four-momentum can be written in terms of energy and momentum as
\begin{eqnarray}
\textsf{\textbf{p}}=\frac{E}{c} {\bf e}_0+{\bf p}' .
\end{eqnarray}
The equation for the norm can then be  rewritten as
\begin{eqnarray}
E^2=m^2c^4+{\bf p}^2 c^2.
\label{Emcp}
\end{eqnarray}

Equation (\ref{fourmomentum}) can also be expressed in terms of spinors
\begin{eqnarray}
\textsf{\textbf{p}}=mc e^{\frac{\bm \alpha}{2}}e^{\frac{\bm \alpha}{2}}{\bf e}_0=
mc e^{\frac{\bm \alpha}{2}}{\bf e}_0e^{-\frac{\bm \alpha}{2}}
\label{fourp}
\end{eqnarray}
where the negative sign in the exponent comes from the anticommutation of ${\bf e}_0$ and ${\bm \alpha}$. Multiplying from the left by $e^{-\frac{\bm \alpha}{2}}$ gives
\begin{eqnarray}
e^{-\frac{\bm \alpha}{2}}\textsf{\textbf{p}}=mc e^{\frac{\bm \alpha}{2}}{\bf e}_0
\end{eqnarray}
 Using a shorthand for the spinor $e^{\frac{\bm \alpha}{2}}={\overline c}+{\overline {\bf s}}$ with ${\overline c}=\cosh\frac{\alpha}{2}$ and ${\overline {\bf s}}={\hat {\bf p}} \sinh\frac{\alpha}{2}$, the above equation splits into a scalar and a vector term
\begin{eqnarray}
\left [(E-mc^2 ){\overline c}  -{\bf p} c {\overline {\bf s}}\right ] +\left [ {\bf p} c {\overline c}+(-E-mc^2) {\overline {\bf s}}
\right ]=0.
\end{eqnarray}
The two independent equations can be written in matrix form as
\begin{eqnarray}
\left (
\begin{array}{cc}
E-mc^2 & -{\bf p} c \\
{\bf p} c  & -E-mc^2
\end{array}
\right )
\left (
\begin{array}{cc}
{\overline c} \\ {\overline {\bf s}}
\end{array}
\right )=0,
\label{Dirac}
\end{eqnarray}
which is essentially Dirac's equation in antisymmetric matrix form \cite{Dirac}. 

From Eq. (\ref{Dirac}), it is clear that there is another solution. This is to be expected since we are dealing with an effective two-dimensional space spanned by ${\bf e}_0$ and ${\hat {\bf p}}$. For a two-dimensional system, the other solution can be found by multiplying the vector with the  unit (pseudo)vector in the exponent of the rotation, as was done in Section \ref{angmom}.  We can apply the this procedure to the hyperbolic rotation
\begin{eqnarray}
R_{{\bm \alpha}\uparrow}^-={\hat {\bf p}}R_{{\bm \alpha}\uparrow}=R_{{\bm \alpha}\uparrow}{\hat {\bf p}}=\sinh \alpha +\cosh\alpha {\hat {\bf p}}.
\end{eqnarray}
This solution is generally denoted as an anti-particle solution of the Dirac equation.
Although the rotation is different, it produces the same four-momentum when using the spinors or the rotations over half times the angle. Following Eq. (\ref{fourp}),
\begin{eqnarray}
\textsf{\textbf{p}}^-&=&mc R_{\frac{\bm \alpha}{2}\uparrow}^- {\bf e}_0 R_{-\frac{\bm \alpha}{2}\uparrow}^-
=mc R_{\frac{\bm \alpha}{2}\uparrow}^-R_{\frac{\bm \alpha}{2}\uparrow}^- {\bf e}_0 \nonumber \\
&=&mc {\hat {\bf p}}R_{\frac{\bm \alpha}{2}\uparrow}{\hat {\bf p}}R_{\frac{\bm \alpha}{2}\uparrow}{\bf e}_0
=R_{{\bm \alpha}\uparrow}{\bf e}_0=\textsf{\textbf{p}},
\end{eqnarray}
where ${\hat {\bf p}}$ and $R_{\frac{\bm \alpha}{2}\uparrow}$ commute, since ${\bm \alpha}={\hat {\bf p}} \alpha$ and therefore commutes with ${\hat {\bf p}}$. However, operating with the four-momentum on $R_{\frac{\bm \alpha}{2}\uparrow}^-$ produces a different result
\begin{eqnarray}
\textsf{\textbf{p}}R_{\frac{\bm \alpha}{2}\uparrow}^- =mc e^{\frac{\bm \alpha}{2}}{\bf e}_0e^{-\frac{\bm \alpha}{2}}  {\hat {\bf p}} e^{\frac{\bm \alpha}{2}}
=-mc {\hat {\bf p}} e^{\frac{\bm \alpha}{2}}{\bf e}_0=-mc R_{\frac{\bm \alpha}{2}\uparrow}^-{\bf e}_0
\nonumber ,
\end{eqnarray}
whereas $\textsf{\textbf{p}}R_{\frac{\bm \alpha}{2}\uparrow}=mc R_{\frac{\bm \alpha}{2}\uparrow}{\bf e}_0$. This shows that the two half rotations are different eigenstates of the four-momentum.

Since the direction of the momentum is also an effective two-dimensional space, one also has to consider the other solution of Eq. (\ref{unitr}). For ${\hat {\bf p}}(-\frac{1}{2})=-{\hat {\bf p}}(\frac{1}{2})$,
\begin{eqnarray}
R_{\frac{{\bm \alpha}}{2}\downarrow}&=&\cosh \frac{\alpha}{2}-{\hat {\bf p}} (-\textstyle{\frac{1}{2}})\sinh \frac{\alpha}{2}  \nonumber \\
R_{\frac{{\bm \alpha}}{2}\downarrow}^-&=&{\hat {\bf p}} (-\textstyle{\frac{1}{2}})R_{\frac{{\bm \alpha}}{2}\downarrow}=\sinh  \frac{\alpha}{2}-{\hat {\bf p}} (-\textstyle{\frac{1}{2}})\cosh \frac{\alpha}{2}  .
\end{eqnarray}
However, since ${\hat {\bf p}}(-\frac{1}{2})=-{\hat {\bf p}}(\frac{1}{2})$, these rotations are equivalent to $R_{\frac{{\bm \alpha}}{2}\downarrow}$ and $R_{\frac{{\bm \alpha}}{2}\downarrow}^-$, respectively. Therefore, it again gives the same four-momentum $\textsf{\textbf{p}}$. Simply stated, this procedures allows the norm of a vector to be negative. Therefore, the same momentum can be created by have a positive norm and a unit vector in the positive direction or a unit vector in the negative direction with a negative norm: $mc (\cosh \alpha {\bf e}_0-\sinh\alpha \,{\hat {\bf p}} (-\textstyle{\frac{1}{2}}) )$.

There are therefore a total of four different rotations $R^\pm_{\frac{{\bm \alpha}}{2}\sigma}$ to obtain the same four-momentum $\textsf{\textbf{p}}$. This result can also be obtained by calculating the eigenvectors $|\psi\rangle$ of the matrix $\textsf{p}=\sum_\mu \gamma_\mu \textsf{p}^\mu$, where $\gamma_\mu$ with $\mu=0,1,2,3$ are the gamma-matrices and calculating the four-momentac
\begin{eqnarray}
\textsf{\textbf{p}} =\sum_\mu \langle {\overline \psi} |\gamma_\mu |\psi\rangle  {\bf e}^\mu,
\end{eqnarray}
with $\langle {\overline \psi} |=\langle \psi|\gamma_0$. This also gives four identical four-momenta for the different eigenvectors.

\section{Motion in an electromagnetic field}
The four-vector potential can be included in the four-momentum by making the following substitution
\begin{eqnarray}
 \textsf{\textbf{p}}\rightarrow  \textsf{\textbf{p}} -q \textsf{\textbf{A}}.
\label{subA}
\end{eqnarray}
We would now like to understand how the motion of a particle is affected by a displacement in spacetime.  The components of the four-potential are conventionally expressed in terms of the scalar potential $V$ and the vector potential ${\bf A}$:
\begin{eqnarray}
\textsf{\textbf{A}}= \frac{V}{c}{\bf e}_0 + {\bf A}',
\label{fourpot}
\end{eqnarray}
At a particular point in spacetime, we can always take $\textsf{\textbf{A}}=0$, assuming that all effects of the potential up to that point have already been included in the four-momentum. A small displacement then gives
\begin{eqnarray}
  \textsf{\textbf{p}}+ d \textsf{\textbf{p}} -q  \raisebox{2pt}{$\bigtriangledown$} \textsf{\textbf{A}}\cdot d\textsf{\textbf{r}},
\end{eqnarray}
and the change in four-momentum should simply be equal to the change in the four-potential: $d \textsf{\textbf{p}} =q  \raisebox{2pt}{$\bigtriangledown$} \textsf{\textbf{A}}\cdot d\textsf{\textbf{r}}$.
The four-vector is given by
\begin{eqnarray}
\textsf{\textbf{r}}=c t+ {\bf r}' ,~~~{\rm and}~~~ d\textsf{\textbf{r}}=c dt+ d{\bf r}',
\end{eqnarray}
where $t$ is the time and $c$ has been included for dimensional reasons; ${\bf r}'={\bf r}{\bf e}_0$ with ${\bf r}'^2=-r^2$ is a vector in real space with norm $r$. The nabla operator in spacetime is
\begin{eqnarray}
 \raisebox{2pt}{$\bigtriangledown$}
&=&\frac{\partial }{c {\bf e}_0 \partial t} + \frac{\partial }{{\bf e}'_x \partial x}+ \frac{\partial }{{\bf e}'_y\partial y}+  \frac{\partial }{{\bf e}'_z\partial z}
 \\
&=&{\bf e}_0 \frac{\partial }{c\partial t} -{\bf e}'_x  \frac{\partial }{\partial x}-{\bf e}'_y \frac{\partial }{\partial y}-{\bf e}'_z  \frac{\partial }{\partial z}
={\bf e}_0 \frac{\partial }{c\partial t}-\nabla', \nonumber
\label{nabla}
\end{eqnarray}
using $1/{\bf e}'_i=-{\bf e}'_i{\bf e}'_i/{\bf e}'_i=-{\bf e}'_i$.
We now want to absorb the change in four-vector potential into the four-momentum giving
\begin{eqnarray}
 d \textsf{\textbf{p}} = q  \raisebox{2pt}{$\bigtriangledown$} \textsf{\textbf{A}}\cdot d\textsf{\textbf{r}}
=\left ( \frac{1}{c} \frac{\partial}{\partial t} {\bf e}_0 -\nabla'
\right )\left (
\frac{V}{c} {\bf e}_0 +{\bf A}' \right )\cdot d\textsf{\textbf{r}} .
\nonumber
\end{eqnarray}
 Expanding the $\raisebox{2pt}{$\bigtriangledown$} \textsf{\textbf{A}}$ gives
\begin{eqnarray}
 \raisebox{2pt}{$\bigtriangledown$} \textsf{\textbf{A}}=
\frac{1}{c^2} \frac{\partial V}{\partial t} -\frac{1}{c} \nabla'  V {\bf e}_0 + \frac{1}{c} \frac{\partial ({\bf e}_0{\bf A}')}{\partial t} -\nabla'{\bf A}'
\end{eqnarray}
The product of the nabla operator and the vector potential is
\begin{eqnarray}
-\nabla'{\bf A}'=\nabla{\bf A}= \nabla \cdot {\bf A}+ \nabla \wedge {\bf A}= \nabla \cdot {\bf A}+{\bf i}  \nabla \times {\bf A}, \nonumber
\end{eqnarray}
using Eq. (\ref{vecprod}). Together with  $\nabla' {\bf e}_0=\nabla$ and ${\bf e}_0{\bf A}'=-{\bf A}$, we find
\begin{eqnarray}
 \raisebox{2pt}{$\bigtriangledown$} \textsf{\textbf{A}}= \raisebox{2pt}{$\bigtriangledown$}\cdot \textsf{\textbf{A}} +
 \frac{1}{c}\left (- \nabla  V  - \frac{1}{c} \frac{\partial {\bf A}}{\partial t}\right ) +{\bf i}  \nabla \times {\bf A}
\label{nablaA}
\end{eqnarray}
The first term $ \raisebox{2pt}{$\bigtriangledown$}\cdot \textsf{\textbf{A}}$ is the inner product between  the four-nabla and the four-vector potential and  contains the two scalar terms. It is equivalent to the Lorenz gauge
\begin{eqnarray}
 \raisebox{2pt}{$\bigtriangledown$}\cdot \textsf{\textbf{A}}
=\frac{1}{c^2} \frac{\partial V}{\partial t} +\nabla \cdot {\bf A}\rightarrow 0
\end{eqnarray}
and we can choose this to be zero by taking the appropriate gauge condition. In the remaining vector and pseudovector terms, we recognize the expressions for the electric and magnetic fields in terms of the potential and the vector potential \cite{Lorrain,Griffiths},
\begin{eqnarray}
{\bf E}=- \nabla  V  - \frac{1}{c} \frac{\partial {\bf A}}{\partial t}, ~~~~~
{\bf B}=\nabla \times {\bf A}.
\end{eqnarray}
Since $ \raisebox{2pt}{$\bigtriangledown$} \textsf{\textbf{A}}= \raisebox{2pt}{$\bigtriangledown$}\cdot \textsf{\textbf{A}}+ \raisebox{2pt}{$\bigtriangledown$}\wedge \textsf{\textbf{A}} $, the vector terms in Eq. (\ref{nablaA})  correspond to the wedge product of the four-vector potential $\textsf{\textbf{A}}$ \cite{Doran},
\begin{eqnarray}
  \textsf{\textbf{F}}=\raisebox{2pt}{$\bigtriangledown$}\wedge  \textsf{\textbf{A}} =\frac{\bf E}{c}+ {\bf B}{\bf i}
	=\frac{\bf E}{c}+ {\underline {\bf B}},
	\label{EMtensor}
\end{eqnarray}
using the definition of the pseudovector from Eq. (\ref{pseudovector}).
 $\textsf{\textbf{F}}$ is known as the electromagnetic field tensor. This bivector contains six distinct terms in the Minkowski basis
\begin{eqnarray}
 \textsf{\textbf{F}} =\raisebox{2pt}{$\bigtriangledown$}\wedge \textsf{\textbf{A}} &=&
F_{x0}{\bf i}'_{x0}+F_{y0}{\bf i}'_{y0}+F_{z0}{\bf i}'_{z0} \nonumber \\
&+&F_{yz}{\bf i}'_{yz}+F_{zx}{\bf i}'_{zx}+F_{xy}{\bf i}'_{xy},
\end{eqnarray}
with ${\bf i}'_{i0}={\bf e}'_i {\bf e}_0={\bf e}_i$ and ${\bf i}'_{ij}={\bf e}'_i{\bf e}'_j={\bf e}_j{\bf e}_i={\bf i}_{ji}$. Conventionally, the six bivectors are described in terms of two real-space vectors, namely the electric and magnetic fields. The first three terms have a space-time bivector  ${\bf i}'_{i0}$ with $i=x,y,z$ and therefore correspond to Lorentz boosts. The amplitude is given by the electric field $F_{i0}=E_i/c$. The last three terms correspond to pure spatial rotations given by the bivector ${\bf i}'_{ij}$ with $i\ne j$. The magnitude is the magnetic field $F_{ij}=B_k$ with $k=x,y,z$ for $ij=yz,zx,xy$.

Writing out the four-vectors, {\it e.g.} $\textsf{\textbf{r}}=ct {\bf e}_0+{\bf r}' =(ct +{\bf r}){\bf e}_0$ and using that $d{\bf r}={\bf v}dt$, we obtain
\begin{eqnarray}
\frac{ d \textsf{\textbf{p}}}{dt}{\bf e}_0=\frac{1}{c} \frac{dE}{dt} + \frac{d{\bf p}}{dt} =q\left (  \frac{\bf E}{c}+ {\bf B}{\bf i} \right ) (c +{\bf v}).
\end{eqnarray}
The vector products can be split into an inner and outer product using Eq. (\ref{vecprod}), for example,
\begin{eqnarray}
 {\bf B}{\bf v}= {\bf B}\cdot {\bf v}+ {\bf i} {\bf B}\times{\bf v}.
\end{eqnarray}
Collecting the terms gives
\begin{eqnarray}
\frac{1}{c} \frac{dE}{dt} + \frac{d{\bf p}}{dt}&=&q\left (  \frac{1}{c} {\bf E}\cdot {\bf v} + {\bf E}+{\bf v}\times {\bf B} \right ) \nonumber \\
&& +{\bf i} qc\left (\frac{1}{c}{\bf B}\cdot {\bf v}+   {\bf B}- \frac{1}{c^2}{\bf v}\times {\bf E}\right   ).
\label{eom}
\end{eqnarray}
The scalar terms are the power equation
\begin{eqnarray}
 \frac{dE}{dt} =&q {\bf E}\cdot {\bf v} .
\end{eqnarray}
The vector terms corresponds to the motion of a particle experiencing a Lorentz force
\begin{eqnarray}
 \frac{d{\bf p}}{dt}&=&q{\bf E}+q{\bf v}\times {\bf B} .
\end{eqnarray}
The pseudovector term in Eq. (\ref{eom}) corresponds to the motion of a particle with a magnetic charge. Magnetic charges are not found in nature. However, even if magnetic charges were present they can be removed from Maxwell's equations by a duality transformation.

\section{Maxwell's equations}
The four-potential appearing in Eq. (\ref{subA}) is due to other charged particles. The generation of the potential is comparable to that of  a driven harmonic oscillator
\begin{eqnarray}
\frac{d^2 x(t)}{dt^2}=-\omega_0^2 x(t)+\frac{f(t)}{m}
\end{eqnarray}
where $x$ is the displacement, $f(t)$ the driving force, $m$ the mass, and $\omega_0$ the frequency of the oscillator. After a Fourier transform to frequency space, one obtains
\begin{eqnarray}
x(\omega)=- \frac{F}{m(\omega^2-\omega_0^2)}
\label{xom}
\end{eqnarray}
For this step, the time dependence of the driving force is assumed to be a $\delta$-function, {\it i.e.} $f(t)=F\delta(t)$ or $f(\omega)=F$. The analogous expression for the four-vector potential is
\begin{eqnarray}
\frac{d^2 \textsf{\textbf{A}}_{\bf{q}}(t)}{dt^2}=-\omega_{\bf q}^2\textsf{\textbf{A}}_{\bf{q}}(t)+\frac{\textsf{\textbf{J}}_{\bf{q}}(t)}{\varepsilon_0},
\label{eomA}
\end{eqnarray}
 where $\varepsilon_0$ is the vacuum permittivity, which effectively plays the role of the mass.
 Since there is also a spatial component in the driving force, we need to include a  dependence on the momentum ${\bf q}$ in the Fourier transform of the four-potential $ \textsf{\textbf{A}}_{\bf{q}}$.  The driving force is the four-current
\begin{eqnarray}
 \textsf{\textbf{J}}( \textsf{\textbf{r}})=c \rho ( \textsf{\textbf{r}}){\bf e}_0+{\bf J}' ,
\end{eqnarray}
where $\rho$ is the charge density and ${\bf J}'$ the charge current. Since charge is a conserved quantity,  the current four-vector should satisfy
\begin{eqnarray}
\frac{\partial \rho}{\partial t}=-\nabla \cdot {{\bf J}'} ~~~{\rm or }~~~ \raisebox{2pt}{$\bigtriangledown$}\cdot \textsf{\textbf{J}}=0.
\label{chargecons}
\end{eqnarray}
Equation (\ref{eomA}) only contains a Fourier transformation of the spatial components. An additional transformation of the time component can be made giving the full Fourier transform
\begin{eqnarray}
 \textsf{\textbf{A}}( \textsf{\textbf{q}})=-\frac{\mu_0 \textsf{\textbf{J}} (\textsf{\textbf{q}})}{\textsf{\textbf{q}}^2},
\end{eqnarray}
where where $\mu_0=1/(c^2\varepsilon_0)$ is the vacuum permeability;
$\textsf{\textbf{q}}=(\omega/c){\bf e}_0 +{\bf q}'$ and $\textsf{\textbf{q}}^2=(\omega/c)^2 -{\bf q}^2$. The above equation can be written as the product of two functions
\begin{eqnarray}
 \textsf{\textbf{A}}( \textsf{\textbf{q}})=-G^0(\textsf{\textbf{q}})\textsf{\textbf{J}} (\textsf{\textbf{q}}),
\end{eqnarray}
where
\begin{eqnarray}
 G^0(\textsf{\textbf{q}})=\frac{\mu_0 }{\textsf{\textbf{q}}^2}=\frac{1}{\varepsilon_0(\omega^2-\omega_{\bf q}^2)},
\label{G0}
\end{eqnarray}
 is the free propagator or Green's function of the photon.  We can obtain the following differential equation for the free propagator
\begin{eqnarray}
 -\textsf{\textbf{q}}^2G^0(\textsf{\textbf{q}})=-\mu_0 ~~~\Rightarrow ~~~ \raisebox{2pt}{$\bigtriangledown$}^2G^0(\textsf{\textbf{r}}) =-\mu_0 \delta(\textsf{\textbf{r}}),
\end{eqnarray}
where the Laplacian in space-time is given by
\begin{eqnarray}
 \raisebox{2pt}{$ \bigtriangledown$}^2 &=&
 \frac{\partial^2}{\partial (ct)^2} -\nabla^2 .
\label{fourLaplacian}
\end{eqnarray}
A product in the momentum-energy dual space is equivalent to a convolution in space-time.
The four-vector potential is then the convolution of the free-particle propagator and the four-current
\begin{eqnarray}
 \textsf{\textbf{A}}( \textsf{\textbf{r}})  = -\int d  \textsf{\textbf{r}}' G^0(  \textsf{\textbf{r}}- \textsf{\textbf{r}}')  \textsf{\textbf{J}}( \textsf{\textbf{r}}').
\end{eqnarray}
The four-vector potential itself satisfies the following differential equation
\begin{eqnarray}
 \raisebox{2pt}{$ \bigtriangledown$}^2 \textsf{\textbf{A}}( \textsf{\textbf{r}})  = \mu_0\textsf{\textbf{J}}( \textsf{\textbf{r}}),
\end{eqnarray}
using the properties of the function $\delta(  \textsf{\textbf{r}}- \textsf{\textbf{r}}') $. This result can also be directly obtained from Eq. (\ref{eomA}). Using Eq. (\ref{EMtensor}), this can be written as
\begin{eqnarray}
 \raisebox{2pt}{$ \bigtriangledown$} \textsf{\textbf{F}}( \textsf{\textbf{r}})  = \mu_0\textsf{\textbf{J}}( \textsf{\textbf{r}})
\label{MaxwellF}
\end{eqnarray}
which are Maxwell's equations in space-time form. Using the expression for the electromagnetic tensor in Eq. (\ref{EMtensor}) and the product of the four-nabla and a spatial vector ${\bf a}$
\begin{eqnarray}
\raisebox{2pt}{$\bigtriangledown$}{\bf a}=\frac{{\bf e}_0}{c}\frac{\partial {\bf a}}{\partial t}+
\nabla \cdot {\bf a}\,{\bf e}_0 - \nabla\times {\bf a}\,{\bf i}_0,
\end{eqnarray}
with
\begin{eqnarray}
{\bf i}_0={\bf e}_0{\bf e}_x{\bf e}_y{\bf e}_z={\bf e}_0{\bf i},
\end{eqnarray}
Equation (\ref{MaxwellF}) can be expanded as
\begin{eqnarray}
&&\left [\frac{1}{c}\left (
\nabla \cdot {\bf E}- \frac{\rho}{\varepsilon_0} \right )
+\left ( \nabla\mathopen{\times} {\bf B} -\frac{1}{c^2}\frac{\partial {\bf E}}{\partial t}-\mu_0 {\bf J}
\right )
\right ] {\bf e}_0
 \nonumber \\
&& ~~~~~ +
\left [ \nabla \cdot {\bf B}
-\frac{1}{c} \left (\nabla\mathopen{\times} {\bf E}+\frac{\partial {\bf B}}{\partial t}
\right ) \right ]{\bf i}_0 =0.
\end{eqnarray}
This results in four independent equations that we recognize as Maxwell's equations.

\section{Nonrelativistic limit}
The vector potential can be included in the four-momentum using Eqs. (\ref{subA})  and (\ref{fourpot}). In the nonrelativistic limit, the approximation $mc \sinh{\bm \alpha}\cong mc{\bm \alpha}\cong {\bf p}-q {\bf A}$ can be used. Since energy is only a conserved quantity in static potentials, the time dependence of the potential is removed. The energy component of the four-momentum is $E-U({\bf r})= mc^2 \cosh{\bm \alpha}$, where $U=qV$ is the potential energy. The hyperbolic function can be expanded as
\begin{eqnarray}
 \cosh{\bm \alpha}\cong 1+ \frac{{\bm \alpha}^2}{2}  =1+ \frac{({\bf p}-q {\bf A}({\bf r}))^2}{2m^2 c^2} .
\end{eqnarray}
The energy can then be written as
\begin{eqnarray}
E-U({\bf r})= mc^2 + \frac{ ({\bf p}-q {\bf A}({\bf r}) )^2}{2m}.
\end{eqnarray}
  Defining the nonrelativistic energy as $E_{\rm nr}=E-mc^2$, this results in
\begin{eqnarray}
 \frac{ ({\bf p}-q {\bf A}({\bf r}) )^2}{2m}+U({\bf r})=E_{\rm nr},
\end{eqnarray}
which, when replacing $E$ and ${\bf p}$ by their quantum mechanical operators is the Schr\"odinger equation. If ${\bf p}$ is an operator, it no longer commutes with the vector potential. The kinetic energy term can then be rewritten as
\begin{eqnarray}
 \frac{ ({\bf p}-q {\bf A}({\bf r}) )^2}{2m}=  \frac{ ({\bf p}-q {\bf A}[{\bf r}] )^2}{2m}+\frac{q}{2m} \{{\bf p}{\bf A}({\bf r})\},
\nonumber
\end{eqnarray}
 In the second term on the right-hand side, the square brackets in the vector potential ${\bf A}[{\bf r}] $ indicate that, although this quantity depends on position, it commutes with the momentum, {\it i.e.} ${\bf p}{\bf A}[{\bf r}]\equiv {\bf A}[{\bf r}]{\bf p}$. The noncommuting part, where ${\bf p}$ only works on the vector potential, is given by curly brackets and corresponds to the anomalous Zeeman effect.

\section{Conclusion}
In this paper, it was shown that several fundamental equations in physics can be obtained by starting from  two fundamental unit vectors denoted as ${\bm a}_\sigma^\dagger$ with $\sigma=\downarrow,\uparrow$. Combinations of the fundamental unit vectors can be used to describe commuting unit vectors ${\bm e}_i$ with $i=x,y,z$ in three-dimensional space. Restriction to a single fundamental vector allows us to obtain anticommuting unit vectors ${\bf e}_i$ that are intimately related to the Pauli spin matrices. A hyperbolic rotation can be used to give vectors a finite norm leading to four-vectors and Dirac equation. In physics, we often deal with the dual space partner of the four-vectors: the four-momentum. The assumption that changes in the four-momentum are given by changes in the four-potential leads to the Lorentz force and the electromagnetic field tensor. Maxwell's equations are obtained by viewing the four-potential as a harmonic oscillator driven by the four-current. The nonrelativistic limit leads to the Schr\"odinger equation.

Obviously, many of the results in the paper are familiar. However, the restriction to a purely geometric approach focusing on the properties of vectors has some interesting consequences. For example, spin and particle/antiparticle solutions already appear at this level, indicating that these are simply different ways to build vectors in space and spacetime. Therefore, the spinor part of the wavefunctions arising from the Pauli-Schr\"odinger and Dirac equations are directly related to the properties of vectors and not necessarily a consequence of (relativistic) quantum mechanics. In particular, spin can be viewed as allowing   vectors with a negative norm, such that a particle traveling in the negative momentum direction with a negative norm produces the same four-momentum. This is not a rejection of wave mechanics. In this paper, it is assumed that the four-momentum (or the energy and the momentum) can be given a certain value at a particular point in space and time. In quantum mechanics, momentum and energy are replaced by operators and their values are "stored" in the wavefunction. This aspect is beyond the scope of the current paper which restricts itself to a description in terms of vectors.

\section{Acknowledgments}
This work was supported by the U. S. Department
of Energy (DOE), Office of Basic Energy Sciences, Division of Materials
Sciences and Engineering under Award No. DE-FG02-03ER46097  and NIU Institute for Nanoscience, Engineering, and Technology. Work at Argonne National
Laboratory was supported by the U. S. DOE, Office of Science, Office of
Basic Energy Sciences, under contract No. DE-AC02-06CH11357.

\bibliography{bib}

\end{document}